\newcommand{\xx}{}	
\ifx\xx\undefined
	\newcommand{\xx}{\hl}
\fi

\documentclass[12pt]{article}

\usepackage{soul}
\usepackage{color}
\usepackage{float}
\usepackage{natbib}
\usepackage{graphicx}
\usepackage{amssymb}
\usepackage{authblk}

\title{Probability distributions for Poisson processes with pile-up}
\author{Diego J. R. Sevilla \\ \texttt{sevilla@ifir-conicet.gov.ar}}
\affil{Rosario Institute of Physics \\ Bv. 27 de Febrero 210bis, S2000EZP Rosario, Argentina}
\date{}

\begin{document}
\maketitle





\begin{abstract}
In this paper, two parametric probability distributions capable to
describe the statistics of X-ray photon detection by a CCD are presented.
They are formulated from simple models that account for the pile-up
phenomenon, in which two or more photons are counted as one.
\xx{These models are based on the Poisson process, but they have an extra
parameter which includes all the detailed mechanisms of the pile-up process
that must be fitted to the data statistics simultaneously with the rate parameter}.
The new probability distributions, one for number of counts per time bins
(Poisson-like), and the other for waiting times (exponential-like)
are tested fitting them to statistics of real data,
\xx{and between them through numerical simulations}, and their
results are analyzed and compared. The probability distributions presented
here can be used as background statistical models to derive likelihood
functions for statistical methods in signal analysis.
\end{abstract}

\section{Introduction}

In the last decades, some statistics methods formulated to find signals
in data sets of X-ray astronomy were developed, some of them based
on bayesian statistics. For example, \citet{gre92} developed a method
able to detect and characterize periodical signals of unknown shapes
and periods. On the other hand, \citet{hut05} developed a method
able to detect changing points on piecewise constant signals, which
could, in principle, be used to determine if a source increases abruptly
its luminosity, as while a burst. Most of these methods assume a parametric
distribution as the background statistical model. Usually, in X-ray
astronomy data analysis, the Poisson distribution is used for this
purpose, because the process of photon detection is considered as
a Poisson process, with a time-dependent rate $r(t)$.

Nevertheless, the process of counting photons in a CCD is not a Poisson process.
Some photon counts result lost because of the pile-up effect \citep{arn11}.
Pile-up occurs when two or more photons are collected by the same
or adjacent pixels in the same frame, so they result counted as only
one event of greater energy.

In this paper, two probability distributions that consider pile-up
are introduced. They are obtained independently: one is based on the Poisson distribution,
and the other, on the exponential probability density. To prove their validity,
they are fitted to real data from RX J0720.4-2125, measured by XMM-Newton
EPIC instrument. \xx{Also, they are compared to each other through numerical simulations.
The results show that the proposed probability distributions describe the statistics
of the real data set remarkably well, and they could be used to analyze similar sets
of data typically found in X-ray astronomical observations.}

The distributions presented in this work could be used to reformulate
or improve some of the statistical methods currently used
in X-ray astronomy data analysis but do not account for the pile-up
effect.

\section{\label{sec:Data}X-ray data acquisition and pile-up}

For X-ray telescopes working in a frame mode, the data consist of
a list of counts where every count ideally corresponds to an arriving
photon. These entries have attributes like detection time, photon
energy, detection pattern (or grade), $x$ and $y$ pixel coordinates, etc. In
this paper we do not discuss the processes of data reduction and selection:
we will concentrate our attention only on the arrival time of the
detected events (counts), which are supposed to come from a single sky source.
All the necessary work to prepare the data we suppose was previously
done.

But in practice, some of the arriving photons are not detected by
the instrument, that is, do not generate counts. On the other hand,
some counts could be not provoked by photons: for example, a count
could be provoked by another incoming particle, or even, by an electronic
noise.

In general, the arrival of photons to a telescope is a Poisson
process: such events are totally independent of one another. The incoming
photons are reflected in a focusing system, and then they are collected
by a detector, which in this case is a CCD array. But some photons
are not reflected in mirrors, and some photons do not interact with
CCDs. These two phenomena\xx{, which depend on photon energy,} are completely stochastic, so they do not
break the initial poissonian character of the arriving photons phenomenon,
that is, the interaction of photons with the detector (or photon collection)
is a Poisson process too. These effects can be described by factors
that give the optical and quantum efficiencies \citep{arn11}. The
estimation of the fraction of photons rejected by them is straightforward,
but it is out of the scope of the present work.

When an individual photon interacts with a CCD, it makes an electrically
charged print in one or more pixels \citep{bal99}. The CCD collects
photons for a time called \emph{frame time}, which is a fundamental
time bin imposed by the instrument. When the frame time finishes,
all the pixels are read, and the information of which pixels were
charged and their respective amounts of electrical charge, is saved.
The reading process takes some time in which the image is moved to
the readout register and the collection of photons of sky sources
occur in spread areas of the image, so the corresponding counts are
usually discarded when the data of a particular sky source are extracted.
Then, the reading times can be considered as dead times in which the
instrument is blind (time gaps), and so they do not alter the poissonian
nature of the photon interaction process.

But there are features of CCD that bias the data and break the poissonian
character of the photon counting process: the most important is pile-up.
As we stated out earlier, pile-up occurs when, in the same frame, two
or more photons hit very close pixels, and they are interpreted as
a single event (count). As pile-up is an instrument saturation effect,
it is not a Poisson process, so the statistics of counts cannot be properly
described by the probability distributions for Poisson processes.

\xx{It is important to note that the probability of pile-up depends on the
instrument characteristics and the energies of the collected photons.
Several authors}
\citep{bal99,dav01}
\xx{have studied pile-up making a detailed analysis of the mechanism of grade migration,
which is fundamental if the objective is the spectral analysis or the luminosity
measure of a stellar source, as pile-up provokes an apparent harder spectrum.
But the scope of this work is different: it is to find probability
distributions capable to describe the statistics of counts of astronomical
X-ray data with no distinction between photon energies. The main purpose
of them is to be used as likelihood functions (i.e., as background statistical
models) in statistical methods devoted to time-series signal shape
characterization, which are commonly used to find periodical pulsations,
burst, etc. in stellar light curves. In this way, we will consider all the
details of pile-up mechanism described in one parameter,
which represents the fraction of counts lost by pile-up.

Finally, the models proposed in this work are essentially different to those
that use the one photon approximation, in which it is assumed that not more
than one photon can be measured per frame}
\citep{kui08}\xx{. While this kind of models are devoted
to obtain the probability to measure $k$ photons in $N$ frames (being $k\leq N$),
the models proposed here follow the purpose to obtain the probability to have
$n$ counts in only one frame. On the other hand, the author has no notice of
a previous study of the pile-up on a waiting-time probability distribution.}

\section{\label{sec:simply-distr}Probability distributions}

\subsection{\label{sub:single_distributions}Probability distributions for Poisson
processes}

\xx{A Poisson process is a stochastic process in which events of a certain kind occur in time
independently of one another. In the case analyzed here, such events are the interactions of
X-ray photons from a stellar source with the CCD detector of an X-ray telescope.
A Poisson process is described by one parameter: the rate of events in time ($r$).
There are two parametric probability distributions usually used to describe poissonian
processes: the Poisson and the exponential distributions.}

\subsubsection{Poisson distribution}

The Poisson distribution is defined as
\begin{equation}
P_{r,\Delta t}(n)=\frac{(r\,\Delta t)^{n}\:\mathrm{\mathbf{e}}^{-r\,\Delta t}}{n!}\label{eq:poisson}
\end{equation}
\xx{and gives the probability to have a number of detections $n$ in a time interval
(i.e., a frame time of the instrument) $\Delta t$, if the rate of detections
in time is $r$. It is a discrete distribution, as $n$ can be any
natural number or zero. From now on, $\Delta t$ is chosen as the
time unit, so the Poisson distribution results}
\begin{equation}
P_{r}(n)=\frac{r^{n}\:\mathrm{\mathbf{e}}^{-r}}{n!}\label{eq:poisson_dist}
\end{equation}
\xx{and $r$ will represent the mean number of detections per frame, if the measurement
is repeated a large number of times with identical conditions.}

\subsubsection{Discrete exponential distribution}

The exponential distribution is defined as
\begin{equation}
f_{r}(\tau)=r\,\mathrm{\mathbf{e}}^{-r\,\tau}\label{eq:exponential_dist}
\end{equation}
and it is a probability density distribution, in the sense that
\[
\int_{t_{1}}^{t_{2}}f_{r}(\tau)\: d\tau
\]
gives the probability to have a waiting time of between $t_{1}$ and
$t_{2}$, where waiting time is the elapsed time between two consecutive
detections. As the instrument reads at regular time steps $\Delta t$,
it is necessary to discretize the exponential distribution in order
to use it to analyze a data set. Using again $\Delta t$ as time unit,
the probability to obtain a waiting time of $n\:\Delta t$ (or simply
$n$), where $n$ can be any natural number or zero, results
\begin{equation}
\left\{
\begin{array}{l}
P_{r}(0)=\int_{0}^{1}(1-t)\: r\: e^{-r\, t}\: dt\\
P_{r}(n)=\int_{n-1}^{n}(1-n+t)\: r\: e^{-r\, t}\: dt+\int_{n}^{n+1}(1+n-t)\: r\: e^{-r\, t}\: dt
\end{array}
\right.
\label{eq:disc_exp_dist-1}
\end{equation}
where the factors in parenthesis are the probabilities of a waiting
time of $t$ to be counted as $n$ or $n+1$, being $n<t<n+1$. Equations
(\ref{eq:disc_exp_dist-1}) can be easily integrated
\begin{equation}
\left\{
\begin{array}{l}
P_{r}(0) = 1+\frac{e^{-r}-1}{r}\\
P_{r}(n) = \frac{e^{-(n+1)\, r}\left(e^{r}-1\right)^{2}}{r}
\end{array}
\right.
\label{eq:disc_exp_dist}
\end{equation}
We will call this probability distribution \emph{discrete exponential distribution}.

\subsection{\label{sub:pile-up_distributions}Distributions for poissonian processes
with pile-up}

When two or more photons arrive to the detector in the same frame,
some of them could be collected on the same place on the CCD, and
interpreted as a single count. This effect, which is called pile-up,
provokes less counts than the number of collected photons. In this
section, probability distributions that consider pile-up are deduced.

\subsubsection{\label{sub:Poisson_pile-up}Poisson distribution with pile-up}

If 2 photons are collected by the CCD in the same frame, there are
two possibilities: the second photon can result piled with the first
one, or not. Let us call $\alpha_{1}^{1}$ to the probability of the
first case, in which the result is only one count. The probability
of the second case, where two counts result, is $1-\alpha_{1}^{1}$. 

Similarly, if three photons are collected by the CCD in the same frame,
there are three possibilities: two counts, one count or none can be
lost. Suppose that the second photon results piled with the first
one which, as we have seen before, occurs with a probability of $\alpha_{1}^{1}$.
When the third photon arrives, the probability of pile-up with the
previous count, which was originated by two piled photons, is $\alpha_{1}^{2}$,
and the probability of not, is $1-\alpha_{1}^{2}$. In the first case,
the result is only one count, and in the second, two counts. Now suppose
that the second photon does not result piled with the first one, which
occurs with a probability of $1-\alpha_{1}^{1}$. Then, when the third
photon arrives, it can result piled with one of the others two, or
not. We call $\alpha_{2}^{1}$ to the probability of the first situation,
in which the photon results piled with one of the two previous counts,
each one originated by one photon, so $1-\alpha_{2}^{1}$ is the probability
of the second situation. In the first situation the result is two
counts, and in the second one, three counts. The idea can be easily
generalized for any number.

\begin{figure}
\begin{center}
\includegraphics[scale=0.6]{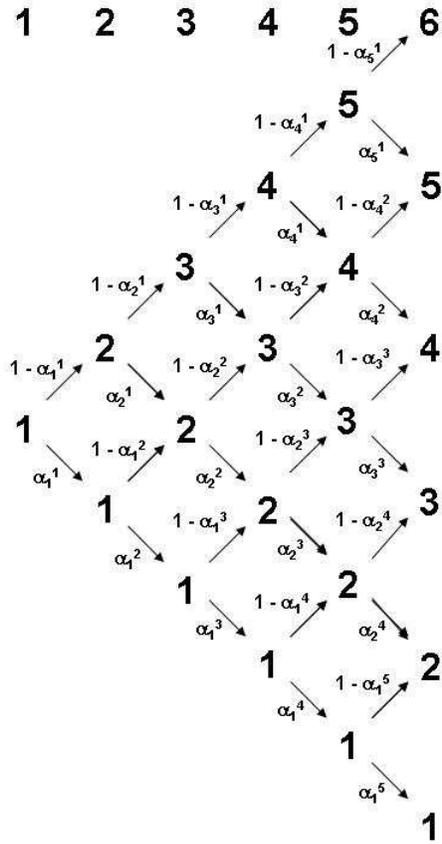}
\end{center}
\caption{\label{fig:tree}Tree of pile-up possibilities. The column numbers
(top numbers) are the numbers of photons collected by the CCD. The
numbers on the nodes are the number of counts that the instrument
records. For each node, there are different possible paths that end
in it. The paths must start in the first node (first column), and
follow the arrows. The probability of the paths can be calculated
multiplying the factors pointed out in the arrows.}
\end{figure}

Figure \ref{fig:tree} shows a tree with the different paths of pile-up
in which we can have $n$ counts from $N$ collected photons, up to
$N=6$. To calculate the probability of a particular path, we must
start from the first column, and pass to the next going to the lower-right
node in case of pile-up, or to the upper-right one if pile-up does
not occur. We repeat this step until column $N$, which represents
the number of photons collected by the CCD. The numbers that label
each node are the numbers $n$ of resulting counts. The probability
of the path can be obtained multiplying the factors that are pointed
out in the arrows that join the nodes. On this way, the sum of the
probabilities of all the possible paths that end in the same column
is 1.

To obtain a formula that expresses the probability of a particular
path is not difficult. Dividing that path in $m$ elementary steps
that have a segment in the upper-right direction, followed by a segment
in the lower-right direction, the probability can be expressed as
\begin{equation}
P_{path}=\prod_{j=0}^{m}\left[\prod_{k=n_{2\, j}+1}^{n_{2\, j+2}}\left(1-\alpha_{k}^{n_{2\, j+1}+1}\right)\:\prod_{k=n_{2\, j+1}+1}^{n_{2\, j+3}}\alpha_{n_{2\, j+2}+1}^{k}\right]\label{eq:Poisson_pile-up_general}
\end{equation}
where $n_{0}=n_{1}=0$, and the increments on the even coefficients
($\Delta n_{j}^{u}=n_{2\, j}-n_{2\, j-2}$) and odd coefficients ($\Delta n_{j}^{d}=n_{2\, j+1}-n_{2\, j-1}$)
are the numbers of nodes that are crossed in the upper-right and the
lower-right directions in the $j^{th}$ step. If the path starts in
the lower-right direction, the first step does not have the upper-right
segment, and $\Delta n_{1}^{u}$ is 0. If the path ends in the upper-right
direction, the last step does not have the segment in the lower-right
direction, and $\Delta n_{m}^{d}$ is 0. The other increments must
be positive numbers.

Equation (\ref{eq:Poisson_pile-up_general}) is general but too complicated
to be useful for pile-up probability calculation. Let us make an assumption
to simplify it: let us suppose that
\begin{equation}
\alpha_{k}^{1}=\alpha_{k}^{2}=.....=\alpha_{k}^{n}=\alpha_{k}\label{eq:assumption_0}
\end{equation}

Assumption (\ref{eq:assumption_0}) considers that the pile-up probability
of a photon that is collected by the CCD only depends on how many
counts are already in the frame, but not on if in these counts some
of the previous collected photons resulted piled. 

To calculate the distribution that gives the probability to obtain
$n$ counts considering that the collected photons follow a Poisson
distribution, we must add the probabilities of all the nodes $n$.
The probability of every node is obtained multiplying the probability
$P_{r}(N)$ for the column of the node (where $N$ is the number of
collected photons and $P_{r}$ is the Poisson distribution (\ref{eq:poisson_dist})),
and the probability obtained adding the probabilities of all the paths
that end in that node. After some combinatorial work, the result is
\begin{equation}
P_{r,\left\{ \alpha_{1},\alpha_{2},...\right\} }(n)=\sum_{m=0}^{\infty}P_{r}(n+m)\,\left(\sum_{k}\,\prod_{j=1}^{m}b_{k\, j}^{(m)}\right)\,\prod_{i=1}^{n-1}\left(1-\alpha_{i}\right)\label{eq:pile-up_poisson_1}
\end{equation}
where $b_{k\, j}^{(m)}$ is the $j^{th}$ element of the m-tuple $B_{k}^{(m)}$
of the set of the m-tuples of $\left\{ \alpha_{1},\alpha_{2},...,\alpha_{n}\right\} $,
considering no permutations. Formula (\ref{eq:pile-up_poisson_1})
is valid for $n\geq1$. On the other hand, $P_{r,\left\{ \alpha_{1},\alpha_{2},...\right\} }(0)$
is equal to $P_{r}(0)$.

The fraction $X$ of lost counts with respect to the total number
of collected photons results
\begin{equation}
X_{r,\left\{ \alpha_{1},\alpha_{2},...\right\} }=\frac{1}{r}\,\sum_{h,k=0}^{\infty}k\, P_{r}(h+k+1)\,\left(\sum_{l}\prod_{j=1}^{k}b_{l\, j}^{(k)}\right)\,\prod_{i=1}^{k}(1-\alpha_{i})
\label{eq:pile_up_poisson_missed_1}
\end{equation}

It is possible to go forward making another assumption on $\left\{ \alpha_{k} \right\}$.
For a low ratio of counts per frame with respect to the total number
of pixels used to acquire the data (i.e. far from saturation),
it is a reasonable assumption to take
\begin{equation}
\alpha_{n}=n\,\alpha\label{eq:assumption}
\end{equation}
This assumption considers that the pile-up probability is proportional
to the number of previous counts in the frame. With it, and after
more combinatorial work, equation (\ref{eq:pile-up_poisson_1}) results
\begin{equation}
P_{r,\alpha}(n)=\sum_{j=1}^{\infty}P_{r}(j)\,\alpha^{j-n}\, S_{j}^{(n)}\,\prod_{k=1}^{n-1}\left(1-k\,\alpha\right)\label{eq:pile-up_poisson_2}
\end{equation}
where $S_{j}^{(n)}$ is the Stirling number of second kind. Again,
formula (\ref{eq:pile-up_poisson_2}) is valid for $n\geq1$. For
$n=0$, the probability is $P_{r}(0)$. In this work, this probability
distribution is called \emph{Poisson distribution with pile-up}, despite
it is not general.

It is to note that probability distribution (\ref{eq:pile-up_poisson_2})
is not a Stuttering-Poisson distribution \citep{gal59}. For a Stuttering-Poisson
distribution, the probability to obtain the case $n$ from the case
$N$ distributed as Poisson, is obtained from the convolution
of $N$ indentically distributed random variables. But for distribution (\ref{eq:pile-up_poisson_2}),
the conditional probability $P_{r,\alpha}(n|N)$ is obtained by adding
the probabilities of all the possible paths of pile-up, being the
probabilities of each step of the paths dependant on the collection history,
that is, on the number of counts that already are in the frame. So,
the process of photon collection with pile-up is not Stuttering-Poisson, and the
probability distribution (\ref{eq:pile-up_poisson_2}) is not a member of the Stuttering-Poisson class.

An important feature of the function that gives the probability to
obtain $n$ counts from $N$ collected photons defined
in equation (\ref{eq:pile-up_poisson_2})
\begin{equation}
P_{r,\alpha}(n|N)=\alpha^{N-n}\, S_{N}^{(n)}\,\prod_{k=1}^{n-1}\left(1-k\,\alpha\right)\label{eq:P_n_N}
\end{equation}
is that it produces alternativelly negative and positive numbers when
$n$ is increased beyond certain value which we call $n^{*}$.
These values do not correspond to physical
probabilities and must be considered as zero, as they are related
to the complete saturation of the instrument. The
last positive value must be corrected in order to satisfy the normalization
condition
\begin{equation}
\sum_{n=0}^{N}P_{r,\alpha}(n|N)=1\label{eq:norma}
\end{equation}

\begin{figure}[H]
\begin{center}
\includegraphics[scale=0.6]{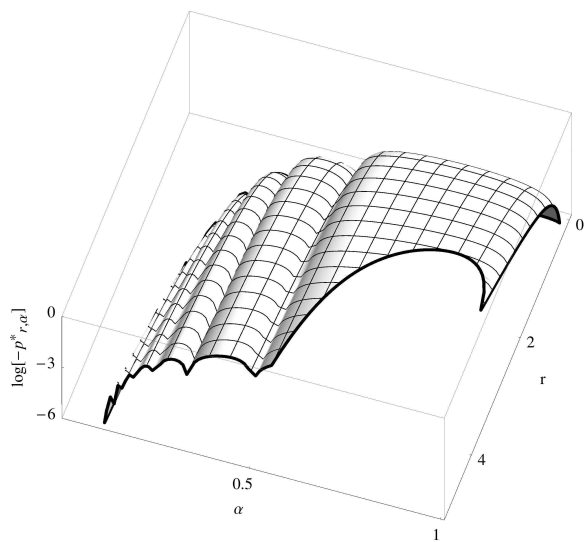}
\end{center}
\caption{
\label{fig:p_asterix}
$\log(-p_{r,\alpha}^{*})$ in function of $r$ and $\alpha$.
This is the first negative value of equation (\ref{eq:pile-up_poisson_2}).
This value and the successive ones must be considered as zero,
and the previous one must be corrected in order to preserve the normalization condition.}
\end{figure}

Figure \ref{fig:p_asterix} \xx{shows the graphics of $\log(-p_{r,\alpha}^{*})$ in function of $r$ and $\alpha$,
where $p_{r,\alpha}^{*}$ is obtained evaluating equation} (\ref{eq:pile-up_poisson_2}) in $n^{*}$.
\xx{As we will see in section} \ref{sec:simulation_tests},
\xx{this procedure is not suitable when} $\left|p_{r,\alpha}^{*}\right|\gtrsim1$
\xx{but, as we can see in figure} \ref{fig:p_asterix}, \xx{it only occurs when $r$ and $\alpha$ are large.}

\xx{Finally, applying assumption} (\ref{eq:assumption}) \xx{to equation} (\ref{eq:pile_up_poisson_missed_1}),
\xx{and considering its Taylor expansion, it can be proved that the fraction $X$ of lost counts with respect to
the total number of collected photons results}
\begin{equation}
X_{r,\alpha}=1+\frac{e^{-\alpha\, r}-1}{\alpha\, r}\label{eq:pile_up_poisson_missed_2}
\end{equation}
\xx{From equation}
(\ref{eq:pile_up_poisson_missed_2})
\xx{we can find the highest possible value of $X$ in function of $r$ taking $\alpha = 1$}
\begin{equation}
X_{max}(r)=1+\frac{e^{-r}-1}{r}\label{eq:Xmax}
\end{equation}

\subsubsection{\label{sub:exponential_pile-up}Discrete exponential distribution
with pile-up}

In this case, when a number $k$ of photons result piled, there is
a missing of $k-1$ waiting times equal to 0 in the data set, while
the numbers of waiting times different to 0 are the same as they would
be if pile-up were null. Calling $X$ to the fraction of lost counts
by the pile-up, and considering equation (\ref{eq:disc_exp_dist}),
the probability distribution of waiting times taking in account the
pile-up results
\begin{equation}
\left\{
\begin{array}{l}
P_{r,X}(0)=\frac{1}{1-X}\,\left(1+\frac{e^{-r}-1}{r}-X\right)\\
P_{r,X}(n)=\frac{1}{1-X}\,\frac{e^{-(n+1)\, r}\left(e^{r}-1\right)^{2}}{r}
\end{array}
\right.
\label{eq:pile-up_discrete}
\end{equation}
We will call \emph{discrete exponential distribution with pile-up}
to probability distribution (\ref{eq:pile-up_discrete}).
\xx{We can see that $P_{r,X}(0)$ is a non negative number only if $X\leq X_{max}(r)$,
where $X_{max}(r)$ is the same previously found for the Poisson distribution with pile-up},
equation (\ref{eq:Xmax}).

\subsection{\label{sub:Error_single_distr}Statistical errors}

Probability distributions like (\ref{eq:pile-up_poisson_2}) or (\ref{eq:pile-up_discrete})
express, if a measure is made, the probability to obtain any of the
different cases labeled as $n$. Probability distributions also express
the fractions of cases $n$ that would be obtained if an infinite
number of measures could be made. But the number of measures that
can be made is finite so, in general, the fractions $q_{n}$ of the
cases $n$ with respect to the total, differ from the probability
distribution $P(n)$, being the differences due to chance. Now the
typical expected differences, which are called statistical errors,
are estimated.

If the probability of happening for the case $n$ in a measure is
$P(n)$, then the probability of not happening is $1-P(n)$. If the
experiment is repeated $N$ times, the probability
to obtain $Q_{n}$ times the case $n$ can be estimated through the binomial
distribution. The most probable number of times in which the event
$n$ occurs is $N\, P(n)$, being the variance $N\, P(n)\,(1-P(n))$.
As $N$ usually is a large number, there is not a significant difference
between the binomial distribution and the normal distribution with
the same mean value and variance. From this, we can estimate the typical
differences between data statistics $q_{n}=Q_{n}/N$ with respect
to their most probable expected values $P(n)$ through the standard
errors of the related normal distributions, that is, the square roots
of the variances
\begin{equation}
\sigma_{n}=\sqrt{\frac{P(n)\,\left(1-P(n)\right)}{N}}\label{eq:simple_error}
\end{equation}

Equation (\ref{eq:simple_error}) is good enough to estimate the statistical
errors for $P(n)>5/N$ \citep{box78}. In that case, it is expected that
$\left|q_{n}-P(n)\right|$ be less than $\sigma_{n}$ for the 68 \%
of the cases, less than $2\,\sigma_{n}$ for the 95\% of the cases,
and less than $3\,\sigma_{n}$ for the 99.7\% of the cases.

\section{\label{sub:pile-up_estimation}Pile-up estimation and rate of photon
collection}

In sections \ref{sub:Poisson_pile-up} and \ref{sub:exponential_pile-up}
two probability distributions able to describe the statistics of data
sets of counts obtained from an intrinsic poissonian process, like
the arrival of photons to an X-ray telescope focal plane, measured
by an instrument which shows pile-up, like a CCD, were deduced. Now,
the problem of how to find the set of parameters that make the best
fit of a probability distribution (statistical model) with respect
to the statistics of a given data set is treated.

Consider the probability distribution $P_{r,s}(n)$ that could be
(\ref{eq:pile-up_poisson_2}) or (\ref{eq:pile-up_discrete}), so
$s$ represents $\alpha$ or $X$. To find the best fitting parameters,
the least squares method can be used. This method consists of finding
the values of the parameters that minimize the quantity
\begin{equation}
\chi^{2}(r,s)=\sum_{n=0}^{n_{max}}\frac{\left(P_{r,s}(n)-q_{n}\right)^{2}}{\sigma_{n}^{2}}\label{eq:ji_square}
\end{equation}
where $\sigma_{n}$ is given by equation (\ref{eq:simple_error}).
The upper limit of the sum, $n_{max}$, which in theory is infinity,
in practice is a number that satisfies, on one hand, that $q_{n}=0$
for $n>n_{max}$, and on the other, that $P_{r,s}(n_{max}+1)\ll1/N$,
being $N$ the number of elements of the data set.

If the probability distribution used to fit the data statistics
is able to describe them, then the statistics of $\chi^{2}(r,s)$
defined in equation (\ref{eq:ji_square}) must be chi-square
distributed, with $n_{max}-1$ degrees of freedom \citep{wal03}.
This is a consequence of the normally and independently distributed
statistical errors of equation (\ref{eq:simple_error}).

To determine the statistical errors of the coefficients $\mathbf{a}=(r,s)$,
the Taylor expansion of $\chi^{2}(r,s)$ around its minimum is used
\begin{equation}
\chi^{2}(\mathbf{a}+\delta\mathbf{a})=\chi_{min}^{2}+\delta\mathbf{a}^{T}\:\mathbf{H}(\mathbf{a})\:\delta\mathbf{a}\label{eq:ji_square_2}
\end{equation}
where $\mathbf{H}$ is the Hessian of $\chi^{2}$, and its components
are equal to \citep{ric95}
\begin{equation}
H_{j\, k}(\mathbf{a})=\sum_{n=0}^{n_{max}}\frac{1}{\sigma_{n}^{2}}\,\left.\frac{\partial P_{r,s}(n)}{\partial a_{j}}\right|_{\mathbf{a}}\,\left.\frac{\partial P_{r,s}(n)}{\partial a_{k}}\right|_{\mathbf{a}}\label{eq:hessian}
\end{equation}

If $\mathbf{C}(\mathbf{a})=\mathbf{H}^{-1}(\mathbf{a})$, the variance
of the parameters result 
\begin{equation}
\sigma_{r}^{2}=\left|\mathbf{C}(\mathbf{a})_{11}\right|\label{eq:sigma_r}
\end{equation}
\begin{equation}
\sigma_{s}^{2}=\left|\mathbf{C}(\mathbf{a})_{22}\right|\label{eq:sigma_s}
\end{equation}
The square roots of these quantities ($\sigma_{r}$ and $\sigma_{s}$)
are the standard errors of the parameters $r$ and $s$ estimated
through the least squares method.

Now, let us consider the effective rate of counts $r_{m}$ measured
by the instrument. It is related to $r$ and $X$ through
\begin{equation}
r_{m}=(1-X)\, r\label{eq:alpha-r-equation}
\end{equation}

Equation (\ref{eq:alpha-r-equation}) is useful to check the consistency
of the results, and it can also be used to improve the estimations
of $r$ and $X$. As (\ref{eq:alpha-r-equation}) permits to obtain
$r(X)$ and $X(r)$, it is possible to promediate $r$ with $r(X)$
and $X$ with $X(r)$, with relative weights according to their statistical
errors \citep{wal03}.
Calling $\hat{r}$ and $\hat{X}$ to these corrected values, they result
\begin{equation}
\frac{\hat{r}}{\sigma_{\hat{r}}^{2}}=\frac{r}{\sigma_{r}^{2}}+\frac{(1-X)^{3}}{r_{m}}\,\frac{1}{\sigma_{X}^{2}}\label{eq:improve_r}
\end{equation}
\begin{equation}
\frac{\hat{X}}{\sigma_{\hat{r}}^{2}}=\frac{X}{\sigma_{X}^{2}}+\left(\frac{r^{4}}{r_{m}^{2}}-\frac{r^{3}}{r_{m}}\right)\,\frac{1}{\sigma_{r}^{2}}\label{eq:improve_alpha}
\end{equation}
where $\sigma_{\hat{r}}$ and $\sigma_{\hat{X}}$ are the standard
errors of $\hat{r}$ and $\hat{X}$
\begin{equation}
\frac{1}{\sigma_{\hat{r}}^{2}}=\frac{1}{\sigma_{r}^{2}}+\frac{(1-X)^{4}}{r_{m}^{2}}\,\frac{1}{\sigma_{X}^{2}}\label{eq:error_improve_r}
\end{equation}
\begin{equation}
\frac{1}{\sigma_{\hat{r}}^{2}}=\frac{1}{\sigma_{X}^{2}}+\frac{r^{4}}{r_{m}^{2}}\,\frac{1}{\sigma_{r}^{2}}\label{eq:error_improve_X}
\end{equation}

For the case of the discrete exponential distribution with pile-up
(\ref{eq:pile-up_discrete}), $X$ and $\sigma_{X}$ are obtained
directly from least squares method and equation (\ref{eq:sigma_s}),
because $X$ is one of the parameters of the distribution. But, for
the case of the Poisson distribution with pile-up (\ref{eq:pile-up_poisson_2}),
$X$ is a function of parameters $r$ and $\alpha$, so $\sigma_{X}$
must be obtained propagating the errors $\sigma_{r}$ and $\sigma_{\alpha}$
in the function (\ref{eq:pile_up_poisson_missed_2})
\begin{equation}
\sigma_{X}^{2}=\left(\frac{\partial X_{r,\alpha}}{\partial r}\right)^{2}\,\sigma_{r}^{2}+\left(\frac{\partial X_{r,\alpha}}{\partial\alpha}\right)^{2}\,\sigma_{\alpha}^{2}\label{eq:propagated_error}
\end{equation}

\section{\label{sec:Aplication}Application to real data}

\subsection{RX J0720.4-3125}

\begin{figure}
\begin{center}
\includegraphics[scale=0.6]{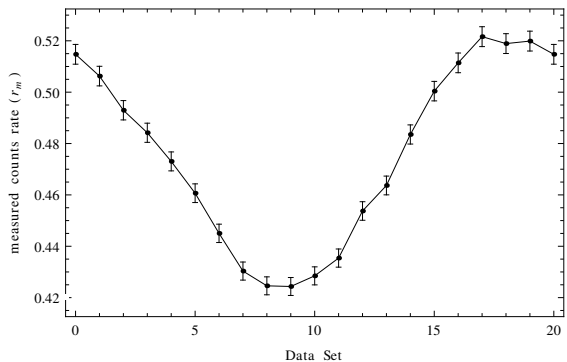}
\end{center}
\caption{\label{fig:light_curve}Mean count rate per time bin ($\sim73$ ms)
measured from RX J0720.4-3125 by EPIC pn instrument at XMM-Newton
Observatory on 2000 May 13. The data were folded in time in 20 bins
per period of $\sim$8.39 s. \xx{The curve is similar to the light curve of the star,
but the abscissas are labeled in data set number instead of star phase.}}
\end{figure}

As application example, a data set from an observation of RX J0720.4-3125
is analyzed. This X-ray source is an isolated and radio-quiet neutron
star. Its luminosity is apparently due to thermical emission, and
it is very regular in an observation-time, except for a periodic change
of about 20\% in its intensity due to its rotation, with a period
of approximately 8.39 s \citep{hoh10}. Also, there is no evidence
of short-time luminosity variations, like sub-second bursts (Sevilla 2013,
in preparation). Figure \ref{fig:light_curve} shows the mean
count rate $r_{m}$ measured by the instrument (see section \ref{sub:Data})
for 20 different star phases, obtained by time folding method \citep{lor05}.
The points are connected by lines for a better visualization.

\subsection{\label{sub:Data}Data}

The data set used was obtained by EPIC pn instrument at XMM-Newton
observatory, operating in full frame mode, with a time resolution
of about 73.4 ms. It corresponds to a continuous observation (without
gaps) of approximately 29.2 ks, extracted from good time intervals
of observation ID 0124100101 on 2000 May 13.

The time bins in data are not regular, because they are expressed
in universal time instead of on-board time, but they can be considered
as equal, as the relative difference between the longest and the shortest
is less than $6\times10^{-4}$, too small to be significant. The cumulative
differences between the universal times of the events and the times
obtained considering a regular time bin result be less than 0.13 time
bin, that is, less than $0.01\:\mathrm{s}$. These differences are
negligible compared to the period of the star, so the time transformation
does not provoke an appreciable phase shift. Finally, only events
of between 120 and 1000 eV are considered. It is because a very low
flux of photons from the source out of this range is expected.

\subsection{\label{sub:folding_distributions}Statistics of data}

As the luminosity of the star is periodic, different subsets of data
close to a given star phase were considered, in order to obtain sets
of data of quasi-constant luminosity. For that, the 10 closest time
bins per period for 20 different and equal spaced phases were selected,
obtaining 20 different data subsets of almost constant luminosities.
Two different lists for every data subset were created: one listing
every time bin and its number of counts, and the other listing the
differences of time between every count and its previous one. Then,
for each list, the fractions $q_{n}$ of cases labeled as $n$ with
respect to the total were calculated. These sets $\{q_{n}\}$ (for
counts per time bin and for waiting times), together with the mean
rate of counts per time bin $r_{m}$, are the statistics needed for
every subset of data. The subsets are sequentially labeled from 1
to 20 (subsets 0 and 20 are the same) and, in all this work, the number
of subset is used to refer to a particular subset instead of its corresponding
star phase.

\begin{figure}
\begin{center}
\includegraphics[scale=0.6]{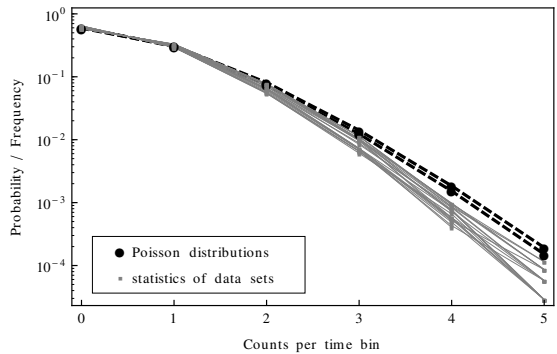}
\end{center}
\caption{\label{fig:statistics-poisson}Comparison of Poisson distributions
(black circles) with the statistics of data for the 20 subsets (gray
squares). For the probability distributions, the highest and the lowest
measured mean rates of counts per time bin $r_{m}$ were considered
(see figure \ref{fig:light_curve}).}
\end{figure}

Figure \ref{fig:statistics-poisson} shows the statistics of measured
counts per time bin for the 20 data subsets (gray squares). It also
shows the Poisson distributions for the highest and the lowest measured
count rates showed in figure \ref{fig:light_curve} (black circles).
Points are connected by lines for a better visualization. These
graphics clearly show that there are less time bins with multiple
counts in the data subsets than the number that the Poisson distributions
for the extreme values of $r_{m}$ predict. 

\begin{figure}
\begin{center}
\includegraphics[scale=0.6]{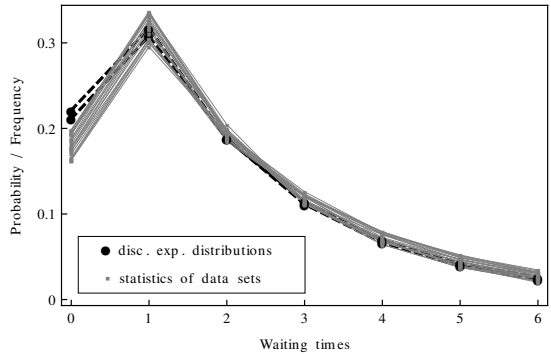}
\end{center}
\caption{\label{fig:statistics-exponential}Comparison of discrete exponential
distributions (black circles) with the statistics of data for the
20 subsets (gray squares). For the probability distributions, the
highest and the lowest measured mean rates of counts per time bin
$r_{m}$ were considered (see figure \ref{fig:light_curve}).}
\end{figure}

Figure \ref{fig:statistics-exponential} shows the statistics of waiting
times for the 20 data subsets (gray squares). It also shows the discrete
exponential distributions for the highest and the lowest measured
count rates showed in figure \ref{fig:light_curve} (black circles).
Again, the points are connected by lines for a better visualization.
These graphics clearly show that there are fewer cases of waiting
time 0 than the number that the discrete exponential distributions
for the extreme values of $r_{m}$ predict.

\subsection{Pile-up estimation and rate of photon collection}

In this section, the probability distributions with pile-up deduced
in section \ref{sub:pile-up_distributions} are fitted to the statistics
of the data subsets, and their parameters are estimated. Poisson and
discrete exponential distributions with pile-up are treated separately,
and finally, their results are compared.

\subsubsection{\label{sub:poisson-fitting}Poisson distribution with pile-up}

Parameters $r$ and $\alpha$ for the best fit of probability distribution
(\ref{eq:pile-up_poisson_2}) to the 20 data subsets obtained in section
\ref{sub:Data} were estimated. It was done calculating $\chi^{2}$
through equation (\ref{eq:ji_square}) for different values of these
parameters, and selecting those that make $\chi^{2}$ minimum. Plots
of $\chi^{2}$ show that it has a quadratic-like behavior in the minimum
neighborhood, so the procedure is justified.

The parameters were considered in the ranges $0.4\leq r\leq0.8$ and
$0.001\leq\alpha<\alpha_{max}$ with steps of $0.001$. The limit
$\alpha_{max}$ was chosen as the minor value of $\alpha$ that makes
$P_{r,\alpha}(n)$ equal to one, for which $\chi^{2}$ has a singular
point, which is related to the complete saturation of the instrument.

\begin{figure}
\begin{center}
\includegraphics[scale=0.6]{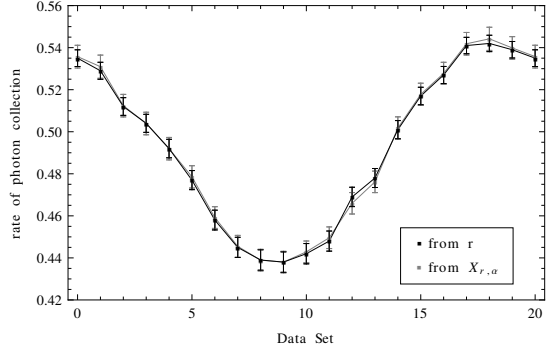}
\end{center}
\caption{\label{fig:r_poisson}Estimation of the rates of photon collection
per time bin for the 20 data subsets fitting the Poisson distribution
with pile-up to the data statistics of counts per time bin. Black
points indicate the estimations using only parameters $r$ and gray
points indicate the estimations using only the values of the fractions
of lost counts $X_{r,\alpha}$.}
\end{figure}

Figure \ref{fig:r_poisson} shows the rate of photon collection per
time bin estimated directly through parameter $r$ (in black), and
from $X_{r,\alpha}$ through equation (\ref{eq:alpha-r-equation})
(in gray), for the 20 different data subsets. The points are joined
by lines for a better visualization. It also shows their respective
error bars. As we can see, both curves are practically coincident,
being the differences consistent with their respective statistical
errors.

\begin{figure}
\begin{center}
\includegraphics[scale=0.6]{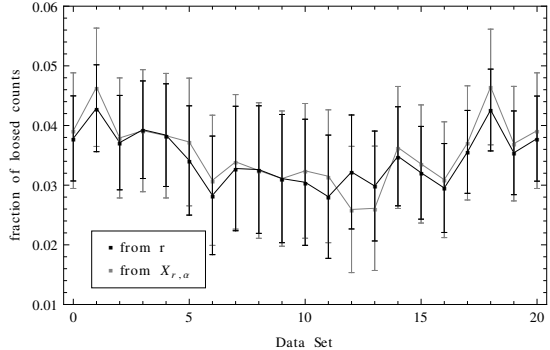}
\end{center}
\caption{\label{fig:X_poisson}Estimation of the fractions of counts lost by
pile-up for the 20 data subsets fitting the Poisson distribution with
pile-up to the data statistics of counts per time bin. Black points
indicate the estimation using only parameter $r$ and gray points
indicate the estimation using only $X_{r,\alpha}$.}
\end{figure}

Figure \ref{fig:X_poisson} shows the fraction of lost counts obtained
from $r$ through equation (\ref{eq:alpha-r-equation}) (in black),
and from $r$ and $\alpha$ through equation (\ref{eq:pile_up_poisson_missed_2})
(in gray), for the 20 data subsets. The points are joined by lines
for a better visualization. It also shows their respective error bars.
Again, we can see that the agreement between both curves is very good,
being the differences consistent with the estimated statistical errors.
It is interesting that the fraction of lost counts seems to have a
variation with respect to the subset number, and then, to the star
phase. This variation is expected, and it is related to the rate of
photon collection: the greater the rate of photon collection, the
greater the fraction of counts lost by the pile-up.

\begin{figure}
\begin{center}
\includegraphics[scale=0.6]{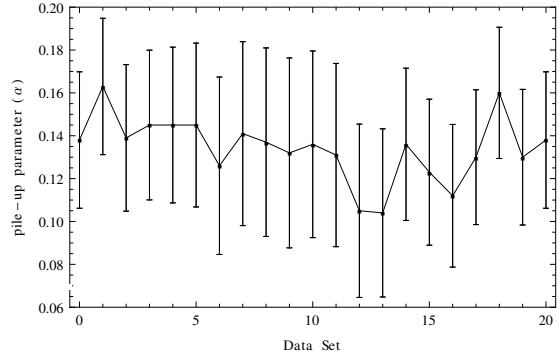}
\end{center}
\caption{\label{fig:alpha_poisson}Estimation of the parameter $\alpha$ that
represents the probability of pile-up of a photon which is collected
from a frame with already one count, for the 20 subsets. The estimations
were made fitting the Poisson distribution with pile-up to the data
statistics of counts per time bin.}
\end{figure}

Figure \ref{fig:alpha_poisson} shows the estimated $\alpha$ for
the 20 data subsets. The points are connected by lines for a better
visualization, and the error bars are shown. It is interesting that
the values of $\alpha$ do not show an appreciable variation that
could be related to the rate of photon collection. Particularly,
around the ninth subset, which corresponds to the minimum rate of
photon collection (figure \ref{fig:r_poisson}), the value of $\alpha$
is close to its mean value for all the subsets. It has a simple
explanation: while $X$ is expected to be dependent on $r$ (the greater
the mean number of photons collected in a frame, the greater the pile-up),
$\alpha$, that expresses the probability of pile-up of a photon with
respect to one previous count in a frame, is not.

\begin{figure}
\begin{center}
\includegraphics[scale=0.6]{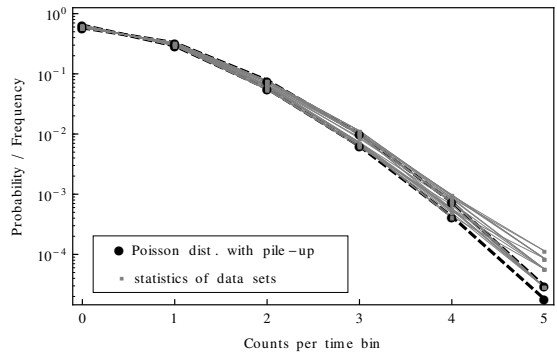}
\end{center}
\caption{\label{fig:poisson_fitting}Comparison of Poisson distributions with
pile-up (black circles), with data statistics of counts per time bin
for the 20 subsets (gray squares). For the probability distributions,
the highest and the lowest values of the parameter $r$ were considered
(figure \ref{fig:r_poisson}), and their corresponding values of $\alpha$
(figure \ref{fig:alpha_poisson}).}
\end{figure}

\begin{figure}
\begin{center}
\includegraphics[scale=0.6]{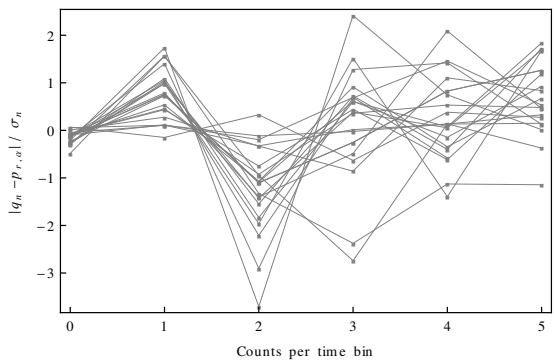}
\end{center}
\caption{\label{fig:Poisson_fitting_error}Differences between the statistics
of counts per time bin and the probabilities given by the Poisson
distributions with pile-up that fit to them, for the 20 data subsets.
The differences are normalized to the standard error $\sigma_{n}$.}
\end{figure}

Figure \ref{fig:poisson_fitting} shows the statistics of measured
counts per time bin for the 20 data subsets (gray squares), and the
Poisson distributions with pile-up for the highest and the lowest
values of parameter $r$ showed in figure \ref{fig:r_poisson}, and
the respective values of parameter $\alpha$ (black circles). Points
are connected by lines for a better visualization. These graphics
show a good agreement between the probability distributions and the
statistics, until 4 counts per time bin. But, the agreement is also
good for 5 counts per time bin: it is only the logarithmic scale in
$y$ axis what magnifies the differences. As the data subsets have
approximately 34,700 time bins, the expected number of bins with 5
counts is only one. In fact, in 8 of the 20 data subsets, there is
only one time bin with 5 counts. The numbers of subsets in which
times bins with 5 counts are present 2, 3 and 4 times are 5, 3 and 2,
respectively. Finally,
there are 2 data subsets with no bins with 5 counts. These last cases
are not shown in figure \ref{fig:Poisson_fitting_error}, because
they correspond to a frequency of $10^{-\infty}$.

Figure \ref{fig:Poisson_fitting_error} shows the differences between
the statistics numbers for the 20 data subsets with respect to the
probabilities given by their corresponding fitting Poisson distributions
with pile-up, normalized to the standard errors ($\sigma_{n}$). It
is evident that the differences are of the order of the statistical
errors for all the cases $n$. Nevertheless, we can see systematic
differences of the order of their standard errors, specially for the
cases of 1 and 2 counts per time bin. But these differences, which
are probably due to the assumptions (\ref{eq:assumption_0}) and (\ref{eq:assumption})
introduced to simplify the probability distribution expression, are
not very important, and the fitting between the probability distributions
and the statistics are quite good.

\begin{figure}
\begin{center}
\includegraphics[scale=0.6]{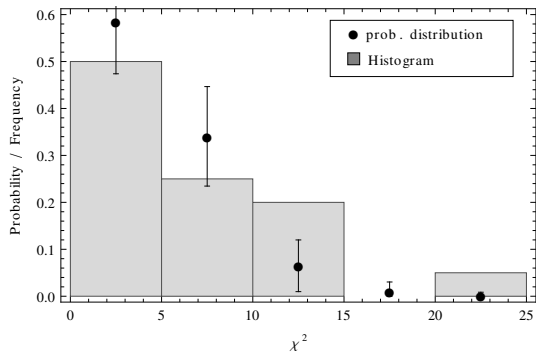}
\end{center}
\caption{\label{fig:histo_poisson}Comparison between the $\chi^{2}(r,\alpha)$
statistics for the 20 data subsets (histogram) and the probable number
of cases for the same intervals calculated with the chi-square distribution
with 5 degrees of freedom (black circles).}
\end{figure}

Finally, figure \ref{fig:histo_poisson} show an histogram of
$\chi^{2}(r,\alpha)$ for the 20 data subsets, and the corresponding
probabilities obtained from the chi-square distribution with 5 degrees
of freedom, with their respective error bars. The width of the histogram
bars was choosen as 5 in order to have some cases in most of the intervals.
The corresponding probabilities were obtained integrating the chi-square
distribution on these intervals. We can see that the agreement
between the statistics of $\chi^{2}(r,\alpha)$ for the 20 subsets,
and the probabilities given by the chi-square distribution, is fair.
The statistics of $\chi^{2}(r,\alpha)$ seem to be more widely dispersed than the values
for the probabilities given by the chi-square distribution. It is due to
the systematic differences stated before (see figure \ref{fig:Poisson_fitting_error}).
Nevertheless, the fitting can be considered as good, so the Poisson distribution with
pile-up appears to be a reasonable statistical model to describe the
statistics of the data subsets.

\subsubsection{\label{sub:exponential-fitting}Discrete exponential distribution
with pile-up}

Again, $\chi^{2}$ was calculated for every data subset, for a range
of $0.2<r<0.8$ and $0.001<X<0.5$, with steps of 0.001, and the values
for the minimum were selected. Again, plots of $\chi^{2}$ show a
good behavior (quadratic-like) that justified the employed method.

\begin{figure}
\begin{center}
\includegraphics[scale=0.6]{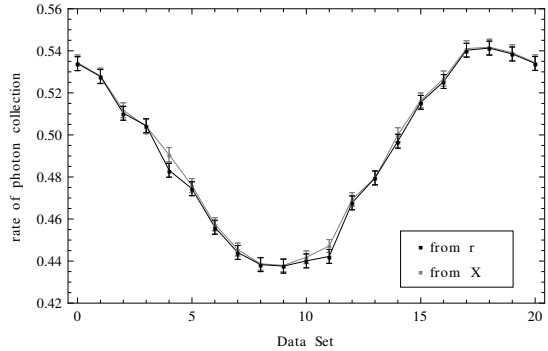}
\end{center}
\caption{\label{fig:corrected_rate}Estimation of the rates of photon collection
per time bin for the 20 data subsets fitting the discrete exponential
distribution with pile-up to the data statistics of waiting times.
Black points indicate the estimations using only parameters $r$ and
gray points indicate the estimations using only the values of the
fractions of lost counts $X$.}
\end{figure}

Figure \ref{fig:corrected_rate} shows $r$ (black points) and $r_{m}/(1-X)$
(gray points) for the 20 different data subsets. The points are joined
with lines, and their respective error bars are also shown. We can
see that the agreement between both quantities is very good, and the
differences are of the order of the estimated statistical errors.

\begin{figure}
\begin{center}
\includegraphics[scale=0.6]{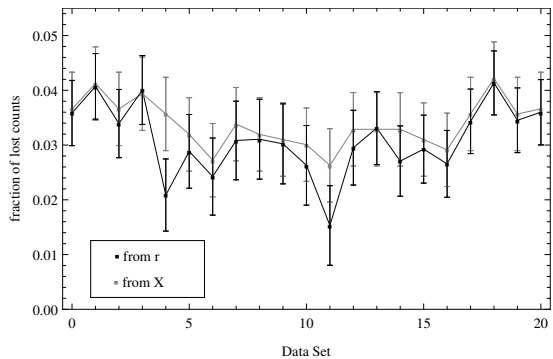}
\end{center}
\caption{\label{fig:missed_counts}Estimation of the fractions of counts lost
by pile-up for the 20 data subsets fitting the discrete exponential
distribution with pile-up to the data statistics of waiting times.
Black points indicate the estimation using only parameter $r$ and
gray points indicate the estimation using only $X$.}
\end{figure}

Figure \ref{fig:missed_counts} shows $X$ (gray points) and $1-r_{m}/r$
(black points), with their respective error bars, for the 20 different
data subsets. The points are joined with lines. Again, we can see
that the agreement between both quantitites is quite good, compatible
with the estimated statistical errors. As in figure \ref{fig:X_poisson},
$X$ shows a variation for the different data subsets, according to
the value of $r$. As it was explained in the previous section, this
variation is expected and has a physical explanation. 

\begin{figure}
\begin{center}
\includegraphics[scale=0.6]{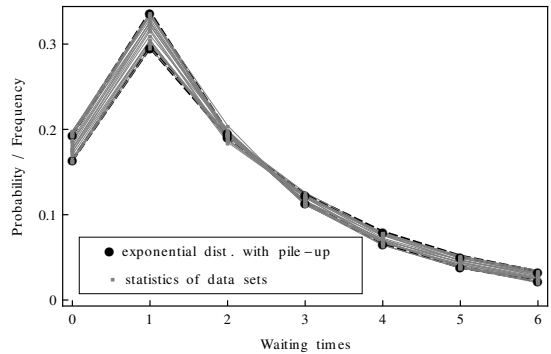}
\end{center}
\caption{\label{fig:exponential_fitting}Comparison of discrete exponential
distributions with pile-up (black circles), with data statistics of
waiting times for the 20 subsets (gray squares). For the probability
distributions, the highest and the lowest values of the parameter
$r$ were considered (figure \ref{fig:corrected_rate}), and their
corresponding values of $X$ (figure \ref{fig:missed_counts}).}
\end{figure}

Figure \ref{fig:exponential_fitting} shows the statistics $q_{n}$
of the measured waiting times for the 20 data subsets (gray squares),
and the discrete exponential distributions with pile-up $P_{r,X}(n)$
for the highest and the lowest values of parameter $r$ showed in
figure \ref{fig:corrected_rate}, and their respective values of $X$
(black circles). Points are connected by lines for a better visualization.
We can see that the agreement between the probability distributions
and the statistics is very good for all the points.

\begin{figure}
\begin{center}
\includegraphics[scale=0.6]{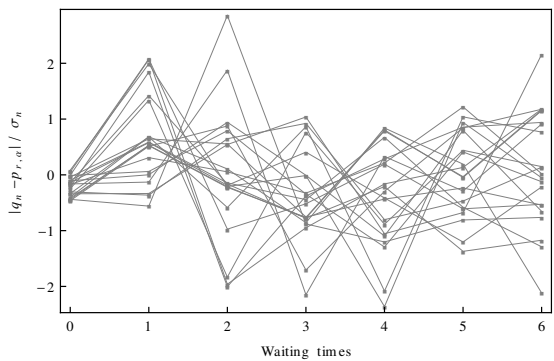}
\end{center}
\caption{\label{fig:exponential_fitting_error}Differences between the statistics
of waiting times and the probabilities given by the discrete exponential
distributions with pile-up that fit to them, for the 20 data subsets.
The differences are normalized to the standard error $\sigma_{n}$. }
\end{figure}

Figure \ref{fig:exponential_fitting_error} shows the differences
between the statistics $q_{n}$ for the 20 data subsets with respect
to the probabilities $P_{r,X}(n)$ given by the discrete exponential
distributions with pile-up that best fit to them, normalized to their
respective standard errors $\sigma_{n}$ (\ref{eq:simple_error}).
Again, we can see that the differences are of the order of the statistical
errors. In this case, if there are systematic differences, they are
not evident. The explanation is that, unlike for the Poisson distribution
with pile-up, for the discrete exponential distribution with pile-up
no assumptions are needed to arrive to a final expression depending
on two parameters: expression (\ref{eq:disc_exp_dist}) depends on
two parameters from the beginning.

\begin{figure}
\begin{center}
\includegraphics[scale=0.6]{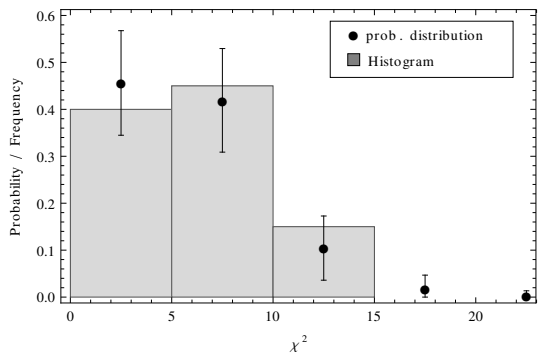}
\end{center}
\caption{\label{fig:histo_exponential}Comparison between the $\chi^{2}(r,X)$
statistics for the 20 data subsets (histogram) and the probable number
of cases for the same intervals calculated with the chi-square distribution
with 6 degrees of freedom (black circles).}
\end{figure}

Finally, figure \ref{fig:histo_exponential} shows an histogram of
$\chi^{2}(r,\alpha)$ for the 20 data subsets, and the corresponding
probabilities obtained from the chi-square distribution with 6 degrees
of freedom. The width of the histogram bars was choosen as 5, and the
corresponding probabilities were obtained integrating the chi-square
distribution on those intervals. Now, we can see that the agreement
between the $\chi^{2}(r,\alpha)$ statistics for the 20 subsets and
the probabilities given by the chi-square distribution is very good. These
results indicate that the discrete exponential distribution with pile-up
is a good statistical model to describe the statistics of the data
subsets.

\subsubsection{Comparison of results}

In sections \ref{sub:poisson-fitting} and \ref{sub:exponential-fitting}
the consistency of the obtained results for each probability distribution
were checked separately. Now, the comparison of the results for both probability
distributions is made.

\begin{figure}
\begin{center}
\includegraphics[scale=0.6]{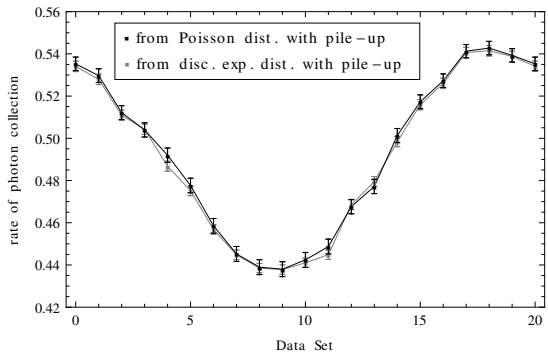}
\end{center}
\caption{\label{fig:comparison_r}Rate of photon collection per time bin obtained
fitting data statistics of counts per time bin to Poisson distribution
with pile-up (black points), and statistics of waiting times to discrete
exponential distribution with pile-up (gray points).}
\end{figure}

Figure \ref{fig:comparison_r} shows the results of the rate of collected
photons obtained by the Poisson distribution with pile-up (black points)
and by the discrete exponential distribution with pile-up (gray points).
The points are joined with lines for a better visualization. The values
were obtained through equation (\ref{eq:improve_r}), which gives
an improved value of them combining the results for the two parameters
of each distribution. Error bars are also shown

We can see that the agreement of the mean value of collected photons
per time bin calculated from both probability distributions is very
good, and the small differences they show can be explained by the
statistical errors.

\begin{figure}
\begin{center}
\includegraphics[scale=0.6]{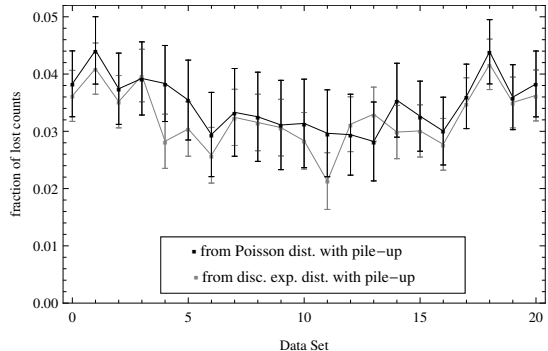}
\end{center}
\caption{\label{fig:comparison_X}Fraction of lost counts obtained fitting
data statistics of counts per time bin to Poisson distribution with
pile-up (black points), and statistics of waiting times to discrete
exponential distribution with pile-up (gray points).}
\end{figure}

Finally, figure \ref{fig:comparison_X} shows the results of the fraction
of lost counts obtained by the Poisson distribution with pile-up (black
points) and the discrete exponential distribution with pile-up (gray
points). The points are joined by lines for a better visualization.
The used values were obtained from equation (\ref{eq:improve_alpha}).
Error bars are also shown. Again, we can see a very good agreement
between the values obtained from both distributions.

\section{\label{sec:simulation_tests}Tests with simulated data}

\xx{In the previous section, the probability distributions proposed
in this work were used to characterize the statistics of a set of real X-ray
astronomical data. These two probability distributions, which arise
from completely different models, were fitted to the statistics
of data, allowing to obtain the rate of photon collection and the fraction
of lost counts, being the results for both probability distributions consistent.
In this section, the two probability distributions
are compared to each other for different values of parameters through
numerical simulations. The aim is to explore when the probability distributions
show consistent results.}

\subsection{Method}

\xx{First, it is important to note that tests with real data can be more decisive than
tests with simulated data, because to simulate data it is necessary to use a model.
If the model used for simulations were related (i.e. has similar assumptions)
to the model used in the method to analyze, the results of the tests could be
a non appropriate validation. Fortunately, in this work, two different
probability distributions, which arise from completely independent models, are presented,
so it is possible to confront them with no danger of obtaining false positive results.
The two probability distributions were confronted as follows: sets
of data obtained through numerical simulations using one probability
distribution (say A), were analyzed with the other one (say B),
in order to see the capability of probability distribution B to
fit on the corresponding statistics of data sets stochastically distributed
as probability distribution A. As the statistical errors of the procedure depend not only
on fittings but also on simulations, for the estimation of the quality of the fittings
of B in data sets generated with A, fittings of probability distribution A on
the same data sets were made. In this way, the capability of probability distribution B to fit on the
corresponding statistics of the data sets can be seen by comparison.} 

\xx{To simulate a set of data using a probability distribution,
first, this probability distribution is truncated in its last relevant
value $p_{M}$. The criterion to determine this value is $p_{M}\lesssim1/N$,
where $N$ is the number of elements of the data set that will be
generated. Then, a random number $x$ between 0 and 1, with more precision
than the magnitude of $p_{M}$, is generated. In this work, $p_{M}\lesssim10^{-4}$,
and the random numbers were generated using the \emph{RandomReal} function
of \emph{Wolfram Mathematica}, with a working precision of 6 (i.e., with
6 digits of precision). Next, the random number $x$ is compared
to $p_{M}$. If it results lower than $p_{M}$, the number $M$
is added to the data set. But if $x$ results greater than $p_{M}$,
then it is compared to $p_{M}+p_{M-1}$, adding the value $M-1$
to the data set if $x$ results lower than that value. If it were
necessary, the same procedure is repeated until one of the numbers
$M$, $M-1$, \ldots{} , 1, 0 results added to the data set, with probabilities
$p_{M}$, $p_{M-1}$, \ldots{} , $p_{1}$, $p_{0}$, respectively.
Then, the data set is transformed into a list of waiting times
(or of counts per time bin, if the simulation were made with the discrete
exponential distribution with pile-up), and the statistics ${q_{n}}$ are calculated
for both lists.}
\xx{The analysis of the data sets with the probability distributions were made
by searching the parameters that make the best fitting of them on the corresponding
statistics of the data sets. The procedure used to find the best fitting
parameters was to minimize the quantity $\chi^{2}$, equation}
(\ref{eq:ji_square}).
\xx{It was done using the function \emph{FindMinimum} of the software \emph{Wolfram Mathematica}.
In this way, from the parameters used in simulations ($r_{sim}$, $\alpha_{sim}$ for the Poisson
distribution with pile-up, and $r_{sim}$, $X_{sim}$ for the discrete exponential distribution with
pile-up), the parameters $r_{fit}$, $\alpha_{fit}$, and $r_{fit}$, $X_{fit}$ that make the best fits
of both probability distributions on the respective statistics of the data sets could be obtained,
and then compared.}

\subsection{Results}

\xx{Several sets of data obtained through the Poisson distribution with pile-up,
with parameters $r_{sim}$, $\alpha_{sim}$ varying between
$r_{min}=0.01$, $r_{max}=5$ and $\alpha_{min}=0.01$, $\alpha_{max}=0.99$ were simulated.
For every choice of parameters, 10 simulations of $10^{4}$ counts were made.
Then, the parameters $r_{fit}$ and $X_{fit}$ that make the best fit of the exponential
distribution with pile-up on the statistics of waiting times were obtained.}

\xx{The parameters obtained by fitting were compared to the ones used
as input in the simulations.
It was done using the folowing parameter}
\begin{equation}
\epsilon_{r}=\frac{r_{sim}-r_{fit}}{r_{sim}-r_{m}}\label{eq:abc}
\end{equation}
\xx{that gives the ratio of the difference between the rates of photon collection
used in simulation and obtained by fitting, with respect to the difference between
the rate of photon collection used in simulation and the statistic $r_m$, that is,
the rate of lost counts. It is possible to make a similar analysis with the parameter
$X$, but the results are quite similar, as both parameters are related trough $r_{m}$
(see equation} (\ref{eq:alpha-r-equation})).
\xx{As the denominator of equation} (\ref{eq:abc}) \xx{is approximately $X \, r_{sim}$ (where $X$ is
$X(r_{sim},\alpha_{sim})$ or $X_{sim}$, depending on which probability distribution
is used on simulation), it is convenient to use $X \, \epsilon_{r}$ for the analysis,
in order to have approximately the same dispersion of values for different parameters.}

\begin{figure}[H]
\begin{center}
\includegraphics[scale=0.6]{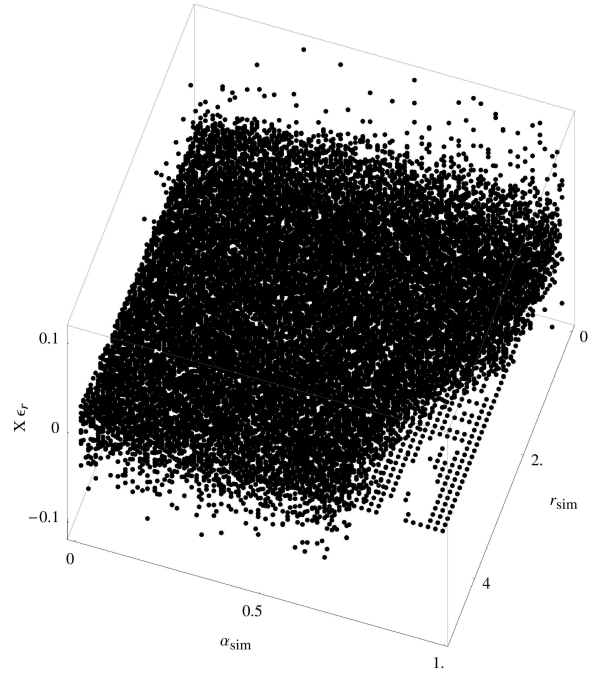}
\end{center}
\caption{
\label{fig:simP_E}
$X\, \epsilon_{r}$ for $r_{fit}$ obtained by fitting the discrete exponential distribution with pile-up
on statistics of data sets obtained by simulation using the Poisson distribution with pile-up,
in function of the simulation parameters $r_{sim}$ and $\alpha_{sim}$.}
\end{figure}

\xx{Figure} \ref{fig:simP_E} \xx{shows the graphics of $X \, \epsilon_{r}$ for the discrete exponential
distribution with pile-up in function of $r_{sim}$ and $\alpha_{sim}$. We can see that the values are close
to 0. For large values of $r$ and $\alpha$, there is a zone in which the fittings usually fail.
In that zone $\left|p^{*}\right|\gtrsim1$ (see figure} \ref{fig:p_asterix}).

\begin{figure}[H]
\begin{center}
\includegraphics[scale=0.6]{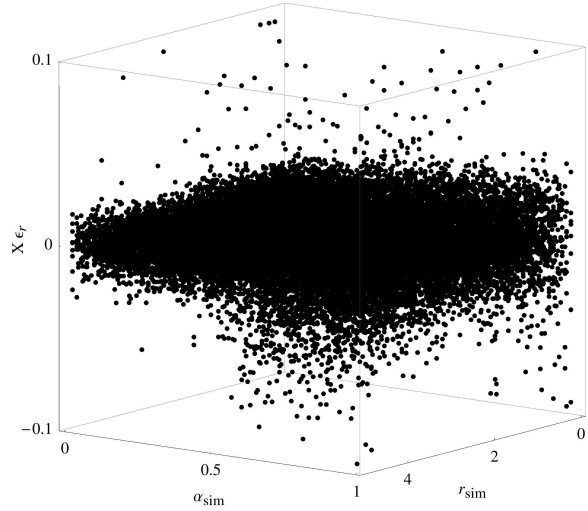}
\end{center}
\caption{
\label{fig:simP_P}
$X\, \epsilon_{r}$ for $r_{fit}$ obtained by fitting the Poisson distribution with pile-up
on statistics of data sets obtained by simulation using the same probability distribution,
in function of the simulation parameters $r_{sim}$ and $\alpha_{sim}$.}
\end{figure}

\begin{figure}[H]
\begin{center}
\includegraphics[scale=0.6]{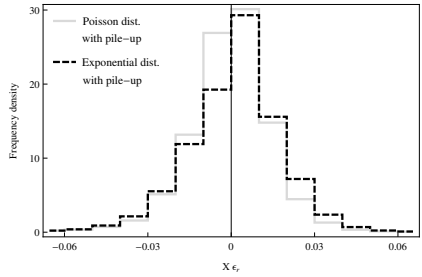}
\end{center}
\caption{
\label{fig:simP_histo}
Distributions of the values of $X\, \epsilon_{r}$ obtained by fitting the discrete exponential distribution
with pile-up (see figure \ref{fig:simP_E}) in black dashed line, and the Poisson distribution with pile-up
(see figure \ref{fig:simP_P}) in gray continuous line, on the corresponding statistics of data sets obtained
by simulation using the Poisson distribution with pile-up.}
\end{figure}

\xx{Figure} \ref{fig:simP_P} \xx{shows the graphics of $X \, \epsilon_{r}$ for the Poisson distribution
with pile-up, the same used in simulations. We can see that the dispersion of values seems similar
to the obtained in the previous case. Figure} \ref{fig:simP_histo}
\xx{shows the distributions of the values of $X \, \epsilon_{r}$ for both fittings.
We can see that the dispersion is quite similar for both cases. It shows that the discrete
exponential distribution with pile-up results an excellent statistical model to describe sets
of data stochastically distributed as the Poisson distribution with pile-up.}

\begin{figure}[H]
\begin{center}
\includegraphics[scale=0.6]{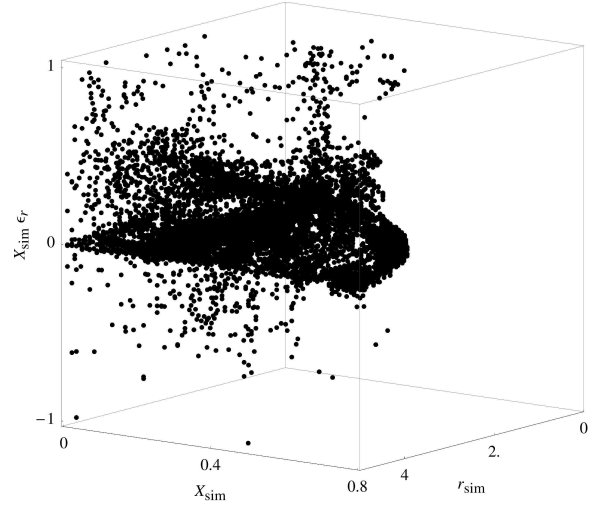}
\end{center}
\caption{
\label{fig:simE_P}
$X\, \epsilon_{r}$ for $r_{fit}$ obtained by fitting the Poisson distribution with pile-up
on statistics of data sets obtained by simulation using the discrete exponential distribution with pile-up,
in function of the simulation parameters $r_{sim}$ and $X_{sim}$.}
\end{figure}

\xx{The same procedure was applied to test the capability of the Poisson distribution with pile-up
to describe the statistics of data sets generated through the exponential distribution with pile-up.
For several pairs of parameters $r_{sim}$, $X_{sim}$ varying between $r_{min}=0.01$, $r_{max}=5$
and $X_{min}=0.01$, $X_{max}=X_{max}(r)$} (equation (\ref{eq:Xmax})),
\xx{10 simulations of $10^{4}$ events were made.
The parameters $r_{fit}$ and $\alpha_{fit}$ that make the best fit of the Poisson distribution
with pile-up on the statistics of counts per frame were obtained. Figure}
\ref{fig:simE_P}
\xx{shows $X_{sim}\, \epsilon_{r}$ in function of $r_{sim}$ and $X_{sim}$. In this case, we can see
that there are several points with large values of $\epsilon_{r}$, and they seem to be not uniformlly
distributed around 0.}

\begin{figure}[H]
\begin{center}
\includegraphics[scale=0.6]{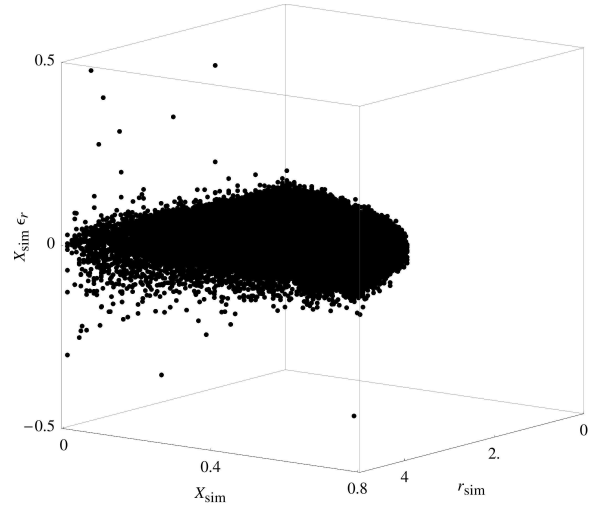}
\end{center}
\caption{
\label{fig:simE_E}
$X\, \epsilon_{r}$ for $r_{fit}$ obtained by fitting the discrete exponential distribution with pile-up
on statistics of data sets obtained by simulation using the same probability distribution,
in function of the simulation parameters $r_{sim}$ and $X_{sim}$.}
\end{figure}

\xx{Figure}
\ref{fig:simE_E}
\xx{shows $X_{sim}\, \epsilon_{r}$ in function of $r_{sim}$ and $X_{sim}$ for the same sets of data,
but now fitted by the discrete exponential distribution with pile-up, which is the same probability
distribution used in simulations. Now we can see that the values of $X_{sim}\, \epsilon_{r}$
are close to 0, and they seem to be uniformlly distributed around this value.}

\begin{figure}[H]
\begin{center}
\includegraphics[scale=0.6]{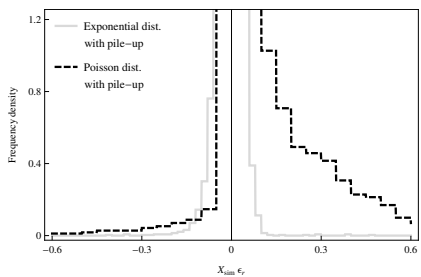}
\end{center}
\caption{
\label{fig:simE_histo}
Distributions of the values of $X\, \epsilon_{r}$ obtained by fitting the Poisson distribution with pile-up
(see figure \ref{fig:simE_P}) in black dashed line, and the discrete exponential distribution with pile-up
(see figure \ref{fig:simE_E}) in gray continuous line, on the corresponding statistics of data sets obtained
by simulation using the discrete exponential distribution with pile-up.}
\end{figure}

\xx{Figure}
\ref{fig:simE_histo}
\xx{shows the distributions of the values of $X_{sim}\, \epsilon_{r}$ obtained by fitting
both probability distributions on the sets of simulated data. In this case we can see a
notorious difference between both distributions. While the values for the discrete exponential
distribution with pile-up seem to have a symetrical distribution around 0, the values for the
Poisson distribution with pile-up present a wider distribution which is not symetrical.}

\xx{Although there are a large number of cases in which the fittings are good, there are a significative
number of situations in which the fittings are not satisfactory. This behaviour points out that
the model of the Poisson distribution with pile-up is more restrictive than the model of the
discrete exponential distribution with pile-up, in the sense that there are sets of data
whose statistics can be described by de second one, but not by the first one.}

\subsection{Conclusions}
\xx{The tests with simulated data show that the Poisson distribution with pile-up and
the discrete exponential distribution with pile-up do not describe different aspects
of the same phenomenon, as the Poisson and Exponential distribution do for a Poisson process,
in the sense that statistics of sets of data stochastically generated with one of them
can be described by the other one and vice versa.
In fact, the discrete exponential distribution with pile-up is able to describe the statistics of a set
of data stochastically generated through the Poisson distribution with pile-up for a wide domain, but the
opposite is not true. It has a simple explanation: while the model used for the formulation of
the discrete exponential distribution with pile-up only assumes
that a fraction of waiting times equal to 0 is lost (which basically is the fundamental
feature of pile-up phenomenon), the Poisson distribution with pile-up
proposed here relies on two arbitrary assumptions}
(equations \ref{eq:assumption_0} and \ref{eq:assumption})
\xx{which are expected to be good aproximations to the behaviour of real detectors
for low values of $r$ and $\alpha$, but they could be replaced by other ones and still satisfying
the model of the discrete exponential distribution with pile-up.
So, the discrete exponential distribution with pile-up corresponds to a more general model, and it
is able to describe the statistics of a whole class of models which includes the corresponding to the
Poisson distribution with pile-up presented in this paper.}

\section{Summary and discussion}

In this paper, two different probability distributions for the description
of statistics of data obtained from measurements of a poissonian process
by an instrument that presents pile-up are deduced. These probability
distributions give the probability to have a number of counts per
time bin, or a certain waiting time between consecutive counts, and
they were deduced from the Poisson and the exponential distributions.
We call these probability distributions Poisson distribution with pile-up
and discrete exponential distribution with pile-up.

A general form of the Poisson distribution with pile-up
is complicated, because it must depend on several parameters,
some of them impossed by the instrument used in data acquisition
\xx{and the photon energy spectrum}.
But, through two assumptions, which are expected to be good approximations
for low rates of photon collection, a simple analytical expression
depending on only two parameters can be obtained (equation \ref{eq:pile-up_poisson_2}),
being this expression independent from instrument features.
These parameters represent the rate of collected photons
per time ($r$), and the probability of pile-up of a collected photon
with a previous count ($\alpha$). But its validity depends on the
validity of the assumptions, so it is not general.

The discrete exponential distribution with pile-up
results very simple. The obtained formula depends on only two parameters:
$r$ (the mean number of collected photons per time bin) and $X$
(the fraction of lost counts) from the beginning, so no extra assumptions
were needed. Then, this distribution results to be general, and so,
valid for any pile-up situation over a poissonian process. In that
sense, the discrete exponential distribution with pile-up is more
simple and robust than the Poisson distribution with pile-up. Nevertheless,
in X-ray data analysis, the Poisson distribution is widely used, but
the exponential distribution, rarely is.

To check the validity of the probability distributions presented here, first,
they were fitted to the same data set using the least squares method,
and their results were analyzed and compared. To do that, a set of real astronomical
data was used. It was obtained from the isolated neutron star RX J0720.4-3125
by EPIC pn instrument at XMM-Newton Observatory. The fittings on the
statistics of data result very good for both distributions, and the
estimations of the mean rates of photon collection, and the fraction
of lost counts, result consistent and practically coincident. So,
it can be concluded that the proposed probability distributions are
valid to describe the statistics of the data of this particular observation,
and it is reasonable to think that they can be valid to describe other
cases similar to this.

\xx{Also, the two probability distributions were tested between them through
numerical simulations. To do that, one of them was used to generate sets
of data for different values of parameters, and the other probability distribution
was fitted to the statistics of the data sets, in order to find the values
of its parameters that describe them best. Then, the values of parameters
obtained fitting the second probability distribution were compared to the
parameters used in the simulation with the first probability distribution,
and the consistency between both models was analyzed.
Both distributions were used in simulations and analysis, and the results
show that discrete exponential distribution with pile-up is able to describe
the statistics of data sets generated by the Poisson distribution with pile-up,
but the opposite is not true.
We conclude that the Poisson distribution with pile-up is less general than the
discrete exponential distribution with pile-up.
This result is consistent with the fact that the model of the Poisson distribution with
pile-up relies on two arbitrary assumptions, but the model of the discrete exponential distribution
with pile-up does not.
So, while the discrete exponential distribution with pile-up seems to be a very general
model able to describe practically any pile-up situation, the validity of the
Poisson distribution with pile-up depends on the accuracy of the assumptions}
(\ref{eq:assumption_0}) and (\ref{eq:assumption}).

\xx{A detailed study of when the Poisson distribution with pile-up presented here is accurate enough
to describe a particular situation, is out of the scope of the present work,
because to determine the pertinence of the assumptions that it uses, it is necessary
to consider in detail the pile-up mechanisms like the grade migration,
which depend on the instrumental characteristics and the photon energy spectrum.}
\xx{But as we saw in section} \ref{sec:Aplication},
\xx{both probability distributions fit
remarkably well on the statistics of the real X-ray astronomical data used in this work,
so we can infer that they must be able to describe the statistics of other similar real data sets.}

Finally, it is to note that the probability distributions proposed
in this paper could be useful to improve some current statistical
methods for X-ray astronomy data analysis that consider the Poisson
distribution as the background statistical model. The Poisson distribution
with pile-up deduced in this work results very simple, and could easily
replace the Poisson distribution in some of these methods.

\xx{\emph{Acknowledgements: I am grateful to my colleagues at Astrophysical Institute
and University Observatory of Friedrich-Schiller-Universitat-Jena,
specially to Valeri Hambaryan, for the support on the observational data used in the present work.
This work was partially supported by the National Council of Scientific and Technical Research
of Argentina (CONICET) and the National University of Rosario.}}


\begin{thebibliography}{}
\bibitem[Arnaud et al.(2011)]{arn11} Arnaud, A. A., Smith, R. K. \& Siemiginowska, A. 2011,
Handbook of X-ray Astronomy (Cambridge: Cambridge Univ. Press)
\bibitem[Ballet(1999)]{bal99} Ballet, J. 1999, A\&AS 135, 371
\bibitem[Box(1978)]{box78} Box, G. E. P., Hunter J. S. \& Hunter, W. G. 1978,
Statistics for experimenters (Wiley)
\bibitem[Davis(2001)]{dav01} Davis, J. E. 2001, ApJ 562, 575
\bibitem[Galliher et al.(1959)]{gal59} Galliher, H. P., Morse, P. M. \& Simond, M. 1959,
Opns. Res. 7, 362
\bibitem[Gregory \& Loredo(1992)]{gre92} Gregory, P. C. \& Loredo T. J. 1992, ApJ 398, 146
\bibitem[Hohle et al.(2010)]{hoh10} Hohle, M. et al. 2010, A\&A 521, A11
\bibitem[Hutter(2005)]{hut05} Hutter, M. 2005, arXiv:math/0606315v1
\bibitem[Kuin(2008)]{kui08} Kuin N. P. M. \& Rosen S. R. 2008, MNRAS 383,383
\bibitem[Lorimer \& Kramer(2005)]{lor05} Lorimer, D. R. \& Kramer, M. 2005,
Handbook of Pulsar Astronomy (Cambridge: Cambridge Univ. Press)
\bibitem[Richter(1995)]{ric95} Richter, P. H. 1995, TDA Progress Report 42-122
\bibitem[Str\"uder(2001)]{str01} Str\"uder, L. et al. 2001, A\&A 365, L18
\bibitem[Wall \& Jenkins(2003)]{wal03} Wall, J. V. \& Jenkins, C. R. 2003,
Practical Statistics for Astronomers (Cambridge: Cambridge Univ. Press)
\end{thebibliography}
\end{document}